\documentclass[aps,prd,preprintnumbers,nofootinbib,twocolumn]{revtex4}
\usepackage{bm}
\usepackage{latexsym}
\usepackage{dcolumn}
\usepackage{amsmath,amsfonts,amssymb}
\usepackage{amsthm}
\usepackage{color}
\usepackage{xcolor}
\usepackage{graphicx,float}
\usepackage{comment}
\usepackage{hyperref}
\usepackage[utf8]{inputenc}
\usepackage[T1]{fontenc}

\def\be {\begin{equation}}
\def\ee {\end{equation}}
\def\bea {\begin{eqnarray}}
\def\eea {\end{eqnarray}}
\def\bc {\begin{center}}
\def\ec {\end{center}}
\def\bfg {\begin{figure}}
\def\efg {\end{figure}}
\def\bi {\begin{itemize}}
\def\ei {\end{itemize}}

\DeclareMathOperator{\Tr}{Tr}

\def\beq{\begin{equation}}
\def\eeq{\end{equation}}
\usepackage{physics}
\def\br{\begin{eqnarray}}
\def\er{\end{eqnarray}}
\newcommand{\eel}[1] {\label{#1}\end{equation}}

\newcommand{\bdm}{\begin{displaymath}}
\newcommand{\edm}{\end{displaymath}}

\begin{document}

\title{Einstein-Rosen bridge from the minimal length}
 
\author{Kimet Jusufi$^\odot$}
\email{kimet.jusufi@unite.edu.mk}

\author{Emmanuel Moulay$^\otimes$}
\email{emmanuel.moulay@univ-poitiers.fr}

\author{Jonas Mureika$^{\square}$}
\email{jmureika@lmu.edu}

\author{Ahmed~Farag Ali$^{\triangle \nabla}$}
\email{aali29@essex.edu ;  ahmed.ali@fsc.bu.edu.eg}

\affiliation{$^\odot$ Physics Department, State University of Tetovo, Ilinden Street nn, 1200, Tetovo, North Macedonia}

\affiliation{$^\otimes$XLIM (UMR CNRS 7252), Universit\'{e} de Poitiers, 11 bd Marie et Pierre Curie, 86073 Poitiers Cedex 9, France}

\affiliation{ $^{\square}$ Department of Physics, Loyola Marymount University, 1 LMU Drive, Los Angeles, CA, USA 90045}

\affiliation{$^\triangle$Essex County College, 303 University Ave, Newark, NJ 07102, United States.}

\affiliation{$^\nabla$Department of Physics, Faculty of Science, Benha University, Benha, 13518, Egypt.}

\begin{abstract}
 We use a string T-duality corrected pair of regular black holes to construct an Einstein-Rosen (ER) bridge with the wormhole throat proportional to the zero-point (Planck) length. This may be a geometric realization of quantum entanglement for particle/antiparticle pairs. We point out that for an extreme mass configuration consisting of a black hole pair, one can have an ER bridge with a horizon area that coincides with the Bekenstein minimal area bound along with a wormhole mass proportional to the Planck mass. This could be related to gravitational self-completeness with quantum mechanical mass limits. We also discuss the ER bridge for  sub-Planckian mass horizonless wormholes and show this admits a region of negative energy at the throat, which we posit to be related to a particle of negative mass generated by quantum fluctuations or the Casimir effect. We argue that Hawking radiation could be the best way for the geometric realization of quantum entanglement for particle/antiparticle pairs emitted by black holes. This sheds new light on the ER=EPR conjecture.
\end{abstract}

\maketitle

\section{Introduction}

In their seminal article \cite{maldacena2013cool}, Maldacena and Susskind proposed a new conjecture linking the Einstein-Rosen (ER) bridges of wormholes \cite{einstein1935particle} with the Einstein–Podolsky–Rosen (EPR) paradox of entangled states \cite{bengtsson2017geometry,einstein1935can}, denoted ER=EPR. In \cite{gao2021traversable,maldacena2016remarks,maldacena2021syk}, the authors developed a traversable Sachdev–Ye–Kitaev (SYK) wormhole in a $(2+1)-$dimensional theory of gravity from \cite{Sachdev:1992fk,kitaev2015simple}. Using a quantum processor for sending quantum information between entangled qubits, this theory has been implemented in \cite{Jafferis2022}. Although this can be viewed as an analog model, it can be an important step in achieving the ER=EPR conjecture in lower-dimensional gravity, whereas the situation is more complex in $(3+1)-$dimensional  gravity \cite{dai2020testing}. Recently, we proved that wormhole geometry contains unitary symmetries in \cite{ali2022unitary} which prompted us to look more closely at the ER=EPR conjecture in the latter context.

It is well known that particle/antiparticle pairs generated in accelerator collisions are entangled \cite{bertlmann2017bell}, as are those created nearby the horizons of black holes \cite{davies1976energy,hawking1975particle}, in vacuum fluctuations \cite{summers2011yet}, or via the Casimir effect \cite{dalvit2011casimir}. The quantum entanglement between a black hole and its Hawking radiation with pair production near the horizon was developed explicitly in \cite[Subsection~3.6]{maldacena2013cool} by using the ER=EPR framework. In this article, we define a new ER bridge, whose throat is proportional to the zero-point length, which may be a geometric realization of quantum entanglement for particle/antiparticle pairs. The associated wormhole metric comes from the black hole metric developed in \cite{Nicolini:2019irw} from string T-duality and bears similarities with the Simpson and Visser metric of traversable wormholes proposed in \cite{simpson2019black}. The resulting wormhole is non-traversable \cite{morris1988wormholes2} since its throat coincides with the minimal length and no particle can have a smaller physical scale \cite{Ali:2009zq,Garay:1994en,Amati:1988tn}. For an extreme mass configuration that consists of a pair of particle-black holes configuration, we can have an ER bridge with an area horizon of $8 \pi l_0^2$ which coincides with the Bekenstein minimal area bound along with a wormhole mass proportional to the Planck mass. What is remarkable with these new kinds of wormholes is that we naturally obtain a region of negative energy at the throat which is related to a particle of negative mass, and we know that this is necessary for the stability of wormholes \cite{hochberg1997geometric,hochberg1998dynamic,ida1999much,morris1988wormholes,morris1988wormholes2,Bambi:2021qfo,Dai:2020rnc,Simonetti:2020ivl,Jusufi:2020yus,Jusufi:2018waj,Jusufi:2021lei} even during inflation where negative energy densities can also generate inflating wormholes \cite{roman1993inflating}. This negative energy necessary for the stability of wormholes responsible for quantum entanglement within the framework of the ER=EPR conjecture could come from quantum fluctuations \cite{kuo1993semiclassical} and be the origin of the dark matter and dark energy in cosmology \cite{farnes2018unifying}. In this article, we would like to study in more detail the possible horizonless wormhole formation in the particle sector by using a string T-duality metric found in \cite{Nicolini:2019irw} in order to tackle the black hole information paradox where Hawking radiation could be the best way for the geometric realization of quantum entanglement for particle/antiparticle pairs emitted by black holes. In this line of thought, we will investigate also the effect of electric charge on the ER bridge. 

The article is organized as follows. We develop new ER bridges with the wormhole throat proportional to the zero-point length having the Planck scale length in Section~\ref{Sec_ERbridge}. These ER bridges are considered with a charge in Section~\ref{Sec_CERbridge}. After showing their embedding diagram in Section~\ref{Sec_Embedding}, we consider Casimir energy as a source of the proposed wormholes in Section~\ref{Sec_Casimir}. The ER=EPR conjecture is analyzed in light of these new wormholes in Section~\ref{Sec_ER=EPR}. Finally, Section~\ref{Sec_Conclusion} concludes the article.

\section{ER bridge in T-duality}\label{Sec_ERbridge}

Consider the static and spherically symmetric metric emerging from string T-duality that solves the Einstein field equations with stringy effect given in \cite{Nicolini:2019irw} by  
\begin{equation}\label{r-metric}
ds^2=-\left(1-\frac{2Mr^2}{\left(r^2+l_0^2\right)^{3/2}}\right)dt^2+\frac{dr^2}{1-\frac{2Mr^2}{\left(r^2+l_0^2\right)^{3/2}}}+r^2 d\Omega^2
\end{equation}
where $d\Omega^2=d\theta^2+\sin^2\theta\,d\varphi^2$ and $l_0$ is the minimal length of Planck order length. Herein, we use a geometrized unit system that sets $G=c=1$. Notice an important feature of the above metric is its invariance under the transformation $r \to -r$. This might already hint that we may describe the whole spacetime with $r \in (-\infty, +\infty)$, and not just focus on the positive region. Later on, we will elaborate on this in more detail. This is a very important solution since it is a non-perturbative one that describes a static and spherically symmetric black hole geometry.  For $M>3 \sqrt{3}\,l_0/4$, there exist two roots:
the inner and outer horizon, $r_-$ and $r_+$, respectively. We can also say that such a metric describes two possible phases of matter, the particle sector when $M<3 \sqrt{3}\,l_0/4$, and the black hole sector when $M>3 \sqrt{3}\,l_0/4$. For a very large mass, the solution is effectively the Schwarzschild black hole. Note that when we compute the surface area of the extreme black hole 
\begin{eqnarray}
    A= \int \sqrt{g} d\theta d\phi= 4 \pi r_{\pm}^2
\end{eqnarray}
with $r_{+}=r_{-}=r^{extr}=\sqrt{2} l_0$, we obtain the interesting result
\begin{eqnarray}
    A_{min}=  8 \pi l_0^2.
\end{eqnarray}
This shows that the black hole area is
quantized in units of $ 8 \pi l_0^2$ and coincides with the standard Bekenstein result \cite{bekenstein1974the}. 
Under the change of coordinates 
\begin{equation}\label{coc}
    u^2=r^2+l_0^2,\,\,\,u_{\pm}= \pm \sqrt{r^2+l_0^2}
\end{equation}
the metric~\eqref{r-metric} becomes
\begin{equation}\label{u-metric}
ds_{\pm}^2=-f(u_{\pm})dt^2+\frac{du_{\pm}^2}{\left(1-\frac{l_0^2}{u_{\pm}^2}\right) f(u_{\pm})}+\left(u_{\pm}^2-l_0^2\right) d\Omega^2
\end{equation}
where 
\begin{equation}
    f(u_{\pm})=1-\frac{2M}{u_{\pm}}+\frac{2M l_0^2}{u_{\pm}^3}.
\end{equation}
If we take the limit $l_0 \to 0$ the wormhole solution reduces to a black hole, namely the Schwarzschild solution. We interpret $u_{\pm}$  mathematically as two congruent parts or ``sheets'', joined by a hyperplane at the wormhole throat. By construction, there should exist a smooth transition between the wormhole interior (a cylinder with length $2 l_0$  whose cross sections are spheres
with the same radius) and the two
external universes.  One can also compute the proper length or distance  traveled as
a function of $u$ which can
easily be found from
\begin{eqnarray}
    l=\pm \int_{u_{min}}^{\pm \infty} \frac{du_{\pm}}{\sqrt{\left(1-\frac{l_0^2}{u_{\pm}^2}\right) f(u_{\pm})}}.
\end{eqnarray}
Solving $f(u_{\pm})=0$ gives the relation
\begin{equation}
    u_{\pm}=\frac{2M}{3}+\frac{4 M^2}{3 \Xi^{1/3}}+\frac{\Xi^{1/3}}{3}
\end{equation}
where 
\begin{equation}
    \Xi=8 M^3-27 M l_0^2+3 \sqrt{81 M^2 l_0^4-48 M^4 l_0^2}.
\end{equation}
We see that $\Xi$ is a real quantity so long as $81 M^2 l_0^4-48 M^4 l_0^2 \geq 0$. From this condition, we obtain the critical mass
\begin{equation}
    M=\pm \frac{3 \sqrt{3} l_0}{4}.
\end{equation}
Note that if we re-write the metric function as 
\begin{equation}
    f(u_{\pm})=1-\frac{2M}{u_{\pm}}\left(1+\frac{ l_0^2}{u_{\pm}^2}\right),
\end{equation}
and substitute the leading order term $u_{\pm} \simeq 2 M+ \mathcal{O}(l_0^2/M^2)$, we get 
\begin{equation}
    f(u_{\pm})=1-\frac{2M}{u_{\pm}}\left(1+\frac{l_0^2}{4 M^2}\right).
\end{equation}
This is the generalized uncertainty principle (GUP) modified energy relation proposed in \cite{Carr:2015nqa} and suggests a new conceptual connection between the GUP and ER bridge. One of the authors \cite{Ali:2022jna} recently found that GUP has the ability to resolve the EPR paradox and offer an explanation of quantum entanglement. This potentially implies that GUP \cite{Amati:1988tn,Ali:2009zq} represents a quantum form of the ER=EPR conjecture that may support the recent experimental result on traversable wormhole dynamics \cite{Jafferis:2022crx}, indicating that the existence of a minimal length is necessary \cite{Ali:2022ckm,Bosso:2022vlz}. 

In this work, we are also interested in the particle domain, hence we need to take 
\begin{equation}
    M < \left|\pm \frac{3 \sqrt{3} l_0}{4}\right|,
\end{equation}
implying that there are no horizons. In such situations, we usually take as physical only the positive solution for the mass. However, we cannot exclude also the possibility of having a negative mass. In fact, the concept of a negative mass was recently used in cosmology to explain dark energy and dark matter \cite{farnes2018unifying}. Later on, we will see that negative mass/energy is needed at the wormhole throat to have an open wormhole. This result is important in the context of the horizon of the metric. We assume that the total spacetime of these two particles is described by the whole spacetime $ds^2$, which has the metric for the  wormhole exteriors $ds_+^2$ and $ ds^2_-$ and the  wormhole interior $
 ds^2_{\rm{wormhole\, interior}} $, which has a cylindric structure and constant radius $r_{throat}=\sqrt{u_{\pm}^2-l_0^2}$. For the exterior coordinates $u_+$, we find the metric function
\begin{equation}
    f(u_+)=1-\frac{2M}{u_+}+\frac{2M l_0^2}{u_+^3},
\end{equation}
and for $u_-=-u_+$, we get 
\begin{equation}
    f(u_-)=1+\frac{2M}{u_+}-\frac{2M l_0^2}{u_+^3}.
\end{equation}
Note the last equation can also be written as 
\begin{equation}
    f(u_-)=1-\frac{2(-M)}{u_+}+\frac{2(-M) l_0^2}{u_+^3},
\end{equation}
which shows that the effect of $u_-$ as seen by an observer located at the first sheet with coordinates $u_+$ is the same as changing the sign before the mass term, i.e. $M \to -M$. This means that it generates a particle with a negative mass. In fact, the ER bridge necessitates the presence of one particle with positive mass and another particle with negative mass. Such regions of
negative energy can be a consequence of quantum fluctuations at short scales \cite{kuo1993semiclassical}. At this point, it is natural to think about taking both solutions and gluing them together to obtain a wormhole which may be a geometric realization of the maximally entangled particle/antiparticle pairs at the wormhole throat, which implies $M=-M$ at the wormhole throat. 
The determinant of this metric~\eqref{u-metric} is 
\begin{eqnarray}
    \det g=u_{\pm}^2>0.
\end{eqnarray}

 To compute the mass we write the metric as 
\begin{equation}
ds_{\pm}^2=-f(u_{\pm})dt^2+\frac{du_{\pm}^2}{1-\frac{2 m(u_{\pm})}{u_{\pm}}}+\left(u_{\pm}^2-l_0^2\right) d\Omega^2
\end{equation}
and the ADM mass of the wormhole is found 
\begin{equation}
   \mathcal{M}=\lim_{u_{\pm} \to \pm \infty } m(u_{\pm})=M.
\end{equation}

The above metric~\eqref{u-metric}, different from the original ER bridge in GR \cite{einstein1935particle}, is not a vacuum solution but it has a source of matter. To see this, let us use the Einstein field equations 
\begin{equation}
    G_{\mu \nu}=8 \pi T_{\mu \nu}
\end{equation}
with
\begin{eqnarray}\label{stress-energy}
    \rho=-P_r=\frac{3 M l_0^2}{4 \pi u^5},\quad P_{\theta}=P_{\phi}=\frac{3 l_0^2 M\left(3 u^2-5 l_0^2\right)}{8 \pi u^7}
\end{eqnarray}
which are regular at $u=l_0$ or $r=0$, meaning $u$ cannot be zero. The energy density describes the density of the elementary particle itself. Assuming that the mass of the particle is $M \sim l_0$, the density of the particle goes like $\rho \sim l_0^{-2}$. In fact, we can write from~\eqref{stress-energy}
\begin{eqnarray}\label{energy_density}
    \rho= \frac{M }{\frac{4}{3}\pi u^3}\frac{l_0^2}{u^2},
\end{eqnarray}
where $4 \pi u^3/3$ plays the role of the effective volume of the particle. This means that when $u=l_0$ we get for the mass of the wormhole the mass of the particle which is consistent with the ER=EPR conjecture in \cite{maldacena2013cool}. Thus, the two sheets of the wormhole with the metrics $ds_+^2$ and $ds_-^2$ may represent the spacetime of entangled particles and antiparticles respectively.

\begin{itemize}
    \item Case $r_{min}=0$
\end{itemize}
When $r=0$, we get $u_{min}= \pm l_0$, hence the interval $u\in \left(-\infty, -l_0\right] \cup \left[l_0, \infty\right)$.  For $u=\pm l_0$, we get $g_{tt}=-f(u)=1$ but there is a coordinate singularity for $g_{rr}$ term.  For the particle sector $M<3 \sqrt{3}\,l_0/4$, the above metric~\eqref{u-metric} describes an ER-like wormhole with a wormhole throat located at $r=0$. It seems at $u=l_0$ there is some sort of coordinate singularity. In principle, this wormhole should not be traversable since its throat coincides with the minimal length and no particle can have scale length $r<l_0$. If we take $r=0$, then the spacetime has a horizon since $u_{min}=l_0$ in that case. To see this we solve $ds^2=d\theta=d\phi=0$, from the metric~\eqref{u-metric} we get 
\begin{equation}
    \frac{du}{dt}= \pm f(u) \sqrt{1-\frac{l_0^2}{u^2}}.
\end{equation}
In particular we have
\begin{equation}\label{throat}
    \left.\frac{du}{dt}\right|_{u_{min}=l_0}= \pm f(u) \sqrt{1-\frac{l_0^2}{u^2}} \to 0.
\end{equation}

The surface area of such a wormhole throat at $u_{min}=\pm l_0$ is zero, i.e. $ A= \int \sqrt{g} d\theta d\phi=0$. It means that such a wormhole is non-traversable for particles. 

\begin{itemize}
    \item Case $r_{min}=l_0$
\end{itemize}
There is another possibility in the metric~\eqref{r-metric} because we have a regular solution for any $r\geq l_0$ since $l_0$ is the minimal length and there is no sense to speak for $r<l_0$. If we set $r_{min}=l_0$ then from Eq.~\eqref{coc} we get the minimal possible value 
\begin{equation}
    u_{min}^2=2 l_0^2,
\end{equation}
which implies that 
\begin{equation}
    u_{min}=\pm \sqrt{2}\, l_0,
\end{equation}
hence $u\in \left(-\infty, - \sqrt{2}l_0\right] \cup \left[ \sqrt{2}l_0, \infty\right)$. 
In terms of $r$, we can now say that the wormhole throat is located at \begin{equation}
    r_{throat}=\sqrt{u_{min}^2-l_0^2}=l_0.
\end{equation}
Since $l_0$ is of the Planck length order, we see that the minimal length plays the role of the wormhole throat. Such a wormhole is not traversable and has no horizon. To see this we solve $ds^2=d\theta=d\phi=0$, from the metric~\eqref{u-metric} we get 
\begin{equation}
    \frac{du}{dt}= \pm f(u) \sqrt{1-\frac{l_0^2}{u^2}},
\end{equation}
which leads to
\begin{equation}
    \left.\frac{du}{dt}\right|_{u_{min}=\sqrt{2}\, l_0}= \pm f(u) \sqrt{1-\frac{l_0^2}{u^2}} \neq 0.
\end{equation}

The surface area of such a wormhole throat at $u_{min}=\sqrt{2} l_0$ is given by
\begin{eqnarray}
    A= \int \sqrt{g} d\theta d\phi= 4 \pi l_0^2.
\end{eqnarray}

The energy density at the wormhole throat $u_{min}=\pm l_0$ gives
\begin{eqnarray}
    \rho|_{u_{min}}=\pm\frac{ 3 M }{4 \pi l_0^3},
\end{eqnarray}
which is equivalent to saying the mass/energy is negative, i.e. $M \to -M$ at the wormhole throat. And for the case $u_{min}=\pm \sqrt{2} l_0$, it gives
\begin{eqnarray}
    \rho|_{u_{min}}=\pm \frac{  3 \sqrt{2} M }{32 \pi l_0^3},
\end{eqnarray}
since we have to match the metric at the wormhole throat, this shows the need for the negative mass/energy at the wormhole. This negative energy at the throat is needed to keep open the wormhole and is consistent with restrictions on negative energy density \cite{roman2006some}. We see that such density depends on the Planck size length and quantum fluctuations can produce such densities in very short scales.

\begin{itemize}
    \item Case $r_{min}=\sqrt{2}l_0$
\end{itemize}
Finally, we study the possibility in the metric~\eqref{r-metric} for $r\geq \sqrt{2} l_0$. From Eq.~\eqref{coc} we get the minimal possible value  
\begin{equation}
    u_{min}=\pm \sqrt{3}\, l_0,
\end{equation}
hence the interval $u\in \left(-\infty, - \sqrt{3}l_0\right] \cup \left[ \sqrt{3}l_0, \infty\right)$.  In terms of $r$, we can now say that the wormhole throat is located at \begin{equation}
    r_{throat}=\sqrt{u_{min}^2-l_0^2}=\sqrt{2} l_0.
\end{equation}

If we solve for $ds^2=d\theta=d\phi=0$, from the metric~\eqref{u-metric} we get 
\begin{equation}
    \left.\frac{du}{dt}\right|_{u_{min}=\sqrt{3}\, l_0}= \pm f(u) \sqrt{1-\frac{l_0^2}{u^2}} \neq 0.
\end{equation}
This means that the spacetime describes a wormhole since there is no horizon provided $M<3 \sqrt{3}\,l_0/4$. Otherwise, for $M=3 \sqrt{3}\,l_0/4$, the spacetime has a horizon and it describes a one-way wormhole. 
The surface area of such a wormhole throat at $u_{min}=\sqrt{3} l_0$ is given by
\begin{eqnarray}
    A= \int \sqrt{g} d\theta d\phi= 8 \pi l_0^2.
\end{eqnarray}

The energy density at the wormhole throat is given by
\begin{eqnarray}
    \rho|_{u_{min}}=\pm \frac{  \sqrt{3} M }{36 \pi l_0^3},
\end{eqnarray}
implying the need for the negative mass/energy at the wormhole throat. 

\section{Charged ER bridge}\label{Sec_CERbridge}

In this section, we consider a charged ER bridge. Toward this goal, we use the charged black hole solution in T-duality given in \cite{Gaete:2022ukm} by  
\begin{equation}\label{r-Cmetric}
ds^2=-f(r)dt^2+\frac{dr^2}{f(r)}+r^2 d\Omega^2
\end{equation}
where
\begin{equation}
    f(r)= 1-\frac{2M r^2}{\left(r^2+l_0^2\right)^{3/2}}+\frac{Q^2 r^2 \mathcal{H}(r)}{\left(r^2+l_0^2\right)^2}
\end{equation} 
with
\begin{eqnarray}
 \mathcal{H}(r)=\frac{5}{8}+\frac{3 l_0^2}{8 r^2}
 -\frac{3\left(r^2+l_0^2\right)^2 }{8 l_0 r^3}\arctan\left(\frac{r}{l_0}\right).
\end{eqnarray}
We can see that when $Q=0$, we get the metric~\eqref{r-metric}.  The ADM mass of the charged wormhole is 
\begin{equation}
   \mathcal{M}=M+\frac{3 \pi Q^2}{16 l_0}.
\end{equation}
Related to the last equation we can ask if a wormhole with zero mass can exist, i.e. $ M=0$, in other words, a wormhole supported by a purely electric. For such a wormhole the mass parameter should be 
\begin{eqnarray}
    \mathcal{M}=\frac{3 \pi Q^2}{16 l_0}.
\end{eqnarray}

Again, in the case of charged ER bridge, $M$ is not the total mass of the system it is only a parameter. As was shown in Ref. \cite{Gaete:2022ukm} the mass is corrected due to the self-electrostatic corrected energy.  In  a similar way, we can perform a change of coordinates 
\begin{equation}
    u^2=r^2+l_0^2,\qquad u= \pm \sqrt{r^2+l_0^2}
\end{equation}
and the metric~\eqref{r-Cmetric} then gives 
\begin{equation}\label{u-Cmetric}
ds_{\pm}^2=-f(u_{\pm})dt^2+\frac{du_{\pm}^2}{\left(1-\frac{l_0^2}{u_{\pm}^2}\right) f(u_{\pm})}+\left(u_{\pm}^2-l_0^2\right) d\Omega^2
\end{equation}
where 
\begin{equation}
    f(u_{\pm})=1-\frac{2M}{u_{\pm}}+\frac{2M l_0^2}{u_{\pm}^3}+\frac{Q^2}{u_{\pm}^2}\left(1-\frac{l_0^2}{u_{\pm}^2}\right)  \mathcal{H}(u_{\pm}),
\end{equation}
with 
\begin{equation}
     \mathcal{H}(u_{\pm})=\frac{5}{8}+\tfrac{3 l_0^2}{8\left(u_{\pm}^2-l_0^2\right)}
 \mp \tfrac{3u_{\pm}^4 }{8 l_0\left(u_{\pm}^2-l_0^2\right)^{3/2}}\arctan\left(\tfrac{\pm \sqrt{u_{\pm}^2-l_0^2}}{l_0}\right).
\end{equation}

The charged solution is also not a vacuum solution but it has a source of matter plus the contribution of the electric field. To see this, let us use the Einstein field equations 
\begin{equation}
    G_{\mu \nu}=8 \pi (T_{\mu \nu}+T^{em}_{\mu \nu}),
\end{equation}
with
\begin{eqnarray}\label{C-stress-energy}
    \rho&=&-P_r=\frac{3 M l_0^2}{4 \pi u^5}+\frac{Q^2(u^2-l_0^2)}{8 \pi u^6},\\\notag
    P_{\theta}&=&P_{\phi}=\frac{3 l_0^2 M\left(3 u^2-5 l_0^2\right)}{8 \pi u^7}+\frac{Q^2 (u^2-l_0^2) (u^2-3 l_0^2) }{8\pi u^8}.
\end{eqnarray}

If we now perform the operation  $u_-=-u_+$, we get 
\begin{equation}
    f(u_-)=1+\frac{2M}{u_+}-\frac{2M l_0^2}{u_+^3}+\frac{Q^2}{u_{+}^2}\left(1-\frac{l_0^2}{u_{+}^2}\right)  \mathcal{H}(u_{+}),
\end{equation}
with $ \mathcal{H}(u_{+})=\mathcal{H}(u_{-})$. The last equation
can also be written as 
\begin{equation}
    f(u_-)=1-\frac{2(-M)}{u_+}+\frac{2(-M) l_0^2}{u_+^3}+\frac{Q^2}{u_{+}^2}\left(1-\frac{l_0^2}{u_{+}^2}\right)  \mathcal{H}(u_{+}) .
\end{equation}
This shows that the effect of $u_-=-u_+$ is the same as changing the sign before the mass term, i.e. $M \to -M$ but the electric charge remains invariant. 

\begin{itemize}
    \item Case $r_{min}=0$
\end{itemize}
For this case, we get $u_{min}=\pm l_0$, but there is a singularity in the $g_{rr}$ term, and the spacetime has a horizon. To see this, again we solve $ds^2=d\theta=d\phi=0$, from the metric~\eqref{u-Cmetric} we get 
\begin{equation}
    \left.\frac{du}{dt}\right|_{u_{min}=l_0}= \pm f(u) \sqrt{1-\frac{l_0^2}{u^2}} \to 0,
\end{equation}
and therefore this spacetime describes a one-way wormhole. The energy density at the wormhole throat $u_{min}=\pm l_0$ gives 
\begin{eqnarray}
    \rho|_{u_{min}}=\pm \frac{  3 M }{4 \pi l_0^3}.
\end{eqnarray}
\begin{itemize}
    \item Case $r_{min}=l_0$
\end{itemize}
In this case, we get $u_{min}=\pm \sqrt{2} l_0$ and the spacetime is everywhere regular. The function 
\begin{equation}
    \left.\frac{du}{dt}\right|_{u_{min}=l_0}= \pm f(u) \sqrt{1-\frac{l_0^2}{u^2}} \neq 0
\end{equation}
since $f(u_{\pm})$ is regular and the wormhole throat is located at $r_{throat}=l_0$.  The energy density at the wormhole throat gives 
\begin{eqnarray}
    \rho|_{u_{min}}=\pm \frac{  3 \sqrt{2} M }{32 \pi l_0^3}+\frac{Q^2}{64 \pi l_0^4}.
\end{eqnarray}
\begin{itemize}
    \item Case $r_{min}=\sqrt{2} l_0$
\end{itemize}
In this case we get $u_{min}=\pm \sqrt{3} l_0$ and the spacetime is everywhere regular
\begin{equation}
    \left.\frac{du}{dt}\right|_{u_{min}=l_0}= \pm f(u) \sqrt{1-\frac{l_0^2}{u^2}} \neq 0
\end{equation}
provided $f(u_{\pm}) \neq 0$.  In the case of charged ER bridge $M$ is not the total mass of the system. Finally, the energy density at the wormhole throat gives 
\begin{eqnarray}
    \rho|_{u_{min}}=\pm \frac{\sqrt{3} M }{36 \pi l_0^3}+\frac{Q^2}{108 \pi l_0^4}.
\end{eqnarray}

\section{Embedding diagram}\label{Sec_Embedding}
\begin{figure}
    \centering
    \includegraphics[scale=0.52]{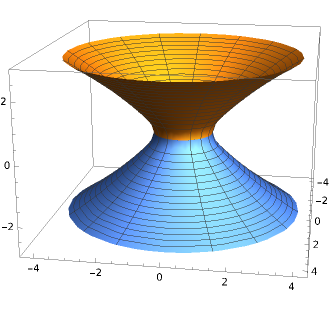}
    \caption{Embedding diagram of the ER bridge with $M=1$, $l_0=1$ and $Q=0$.}
    \label{fig:embedding}
\end{figure}

In this section, we will analyze the embedding diagram that helps us to impose the demand of 
the spacetime metric~(\ref{u-metric}) to describe a wormhole. Of particular interest in the geometry, we 
consider an equatorial slice $\theta=\pi/2$ at some fixed moment in time $t=const$. 
With this constraint the metric~(\ref{u-metric}) becomes,
\begin{equation}\label{u-metric_emb1}
    ds^2=\frac{du^2}{\left(1-\frac{l_0^2}{u^2}\right)f(u)}+\left(u^2-l_0^2\right)d\phi^2.
\end{equation}
The reduced metric~(\ref{u-metric_emb1}) can be embedded into a $3$-dimensional
Euclidean space, and in cylindrical coordinates $r$, $\phi$ and $z$ has the form
\begin{equation}\label{u-metric_emb2}
    ds^2=dz^2+du^2+(u^2-l_0^2)d\phi^2.
\end{equation}
The embedded surface $z(r)$ can be obtained by reversing and integrating 
from the last two equations, we obtain the slope
\begin{equation}
    \frac{dz}{du}=\pm \sqrt{\frac{1}{\left(1-\frac{l_0^2}{u^2}\right)f(u)}-1}.
\end{equation}

With the illustration of Fig.~\ref{fig:embedding}, we explore the geometrical properties of these metrics~(\ref{u-metric_emb1}) and (\ref{u-metric_emb2}) via the embedding diagram. Numerical values are enlisted in the caption of Fig.~\ref{fig:embedding}.

\section{Wormhole with Casimir energy}\label{Sec_Casimir}

In this section, we will try to give further physical arguments about the concrete formation of the wormholes defined above. The first example is to consider Casimir energy as a source of such wormholes which were studied in \cite{Garattini:2019ivd}.  The energy density in such a case is indeed negative [here for a moment we restore the constants $c$ and $\hbar$]
\begin{equation}
\rho_{C}\left(  a\right)  =-\frac{\hbar c\pi^{2}}{720a^{4}}\label{rhoC}%
\end{equation}
where $a$ gives the constant plate separation. 
The total Stress-Energy Tensor (SET) for such a case could be  \cite{Garattini:2019ivd}
\begin{eqnarray}\notag
T_{\sigma}^{\mu\nu}&=&\sigma\hat{t}^{\mu}\hat{t}^{\nu}\left[  \delta\left(
z\right)  +\delta\left(  z-a\right)  \right]  \\
&&+\Theta\left(  z\right)
\Theta\left(  a-z\right)  \frac{\hbar c\pi^{2}}{720a^{4}}\left[  \eta^{\mu\nu
}-4\hat{z}^{\mu}\hat{z}^{\nu}\right]  ,\label{TCas}
\end{eqnarray}
where $\hat{t}^{\mu}$ is a unit time-like vector, $\hat{z}^{\mu}$ is a normal vector to the plates and $\sigma$ is the mass density of the plates. If we combine energy density~\eqref{energy_density} at the throat $u_{min} \sim l_0$ and we further let $a \to u$, we get the mass needed for such a wormhole
\begin{eqnarray}\label{mass}
    M \sim  \frac{\hbar\, \pi^{4}\,l_0^3}{540\,c\, u^{4}}.
\end{eqnarray}
If we further take $u \sim l_0$, and we take $l_0 \sim l_{Pl}=\sqrt{\frac{\hbar c}{G} }$ we obtain 
\begin{equation}\label{WH_mass}
    M \sim \,\frac{\pi^4}{540}\,M_{Pl}
\end{equation}
where $M_{Pl}=\sqrt{\frac{\hbar c}{G}}$ is the Planck mass. We get for such a wormhole mass $M \sim 10^{-9}$~kg. This seems to be a large amount of mass to be generated by quantum fluctuations for entangled particles. If for instance we choose $u \sim 10^{-30}$~m in the expression for the mass~\eqref{mass} we get $M \sim 10^{-25}$~kg.  This simply means that having such an extreme mass in a very short distance like $u \sim l_0$ will lead to the collapse and formation of the particle-black hole candidate. In other words, the Planck mass acts like the upper bound mass for a pair of entangled particle-black holes.  However, for larger distances for $u$, as we saw from the Casimir density, we indeed get a small mass, say $M \sim 10^{-25}$~kg, many orders of magnitude smaller than the Planck mass and thus compatible with the elementary particle mass. In principle, such a mass can be generated by quantum fluctuations for pair of entangled particles even in a vacuum. 

\section{Testing ER=EPR}\label{Sec_ER=EPR}

\begin{figure}
    \centering
    \includegraphics[scale=0.55]{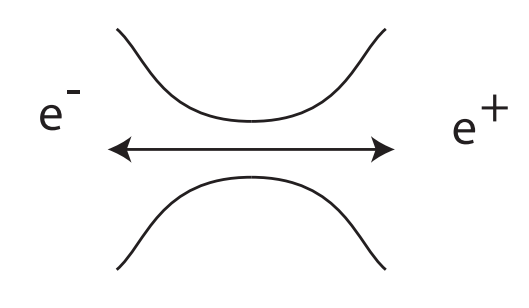}
    \caption{Schematic plot from \cite{dai2020testing} of entangled $e^+ e^-$ pair and the ER bridge connecting the particles.}
    \label{fig:entangled_pair}
\end{figure}

We follow a similar analysis done in \cite{dai2020testing}. In particular, we will consider another example of entangled particles that could lead to a wormhole formation: a pair of particles with extreme mass that can undergo a gravitational collapse into a black hole. Such a spacetime has horizon at $r=\sqrt{2} l_0$ and mass $M=3\sqrt{3}l_0/4$. As an example, we can consider a $e^+ e^-$ pair created in a maximally entangled state connected by an ER bridge with the spin state 
\begin{equation}
\ket {\Psi} = \frac{1}{\sqrt{2}}\left(\ket{\uparrow_1 \downarrow_2}+\ket{\downarrow_1 \uparrow_2}\right),
\end{equation}
see Fig.~\ref{fig:entangled_pair}.
These particle-black holes are entangled. From the first law of black hole thermodynamics, we can find the Bekenstein-Hawking entropy related to the metric~\eqref{r-metric}. The entropy at the horizon reads
\begin{eqnarray}
    S&=&\frac{k_B c^3 A }{ 4 G \hbar}\Big(1-\frac{8 \pi l_0^2}{A}\Big)\sqrt{1+\frac{4 \pi l_0^2}{A}}\\\notag
    &&+ \frac{3 \pi k_B c^3  l_0^2}{ G \hbar} \ln \left[ \frac{1}{2\, \mathcal{C}} \left( \sqrt{\frac{A}{\pi}}+ \sqrt{\frac{A+4\pi l_0^2}{\pi}}\right)\right],
\end{eqnarray}
where $\mathcal{C}$ is an integration constant that we can set to $\mathcal{C}=l_0$. In this section, we use \emph{international standard units system}. As we pointed out, we take $r=\sqrt{2} l_0$,  or $A= 8 \pi l_0^2$, which plays the role  of the cut through the ER bridge with minimal length, this gives 
\begin{eqnarray}\label{Entropy_1}
    S=\frac{3 \pi k_B c^3 l_0^2  }{G \hbar}\ln(\sqrt{2}+\sqrt{3}).
\end{eqnarray}
On the other hand, the entropy must be larger than
the entanglement entropy between the electron and positron in a pair given by
\begin{eqnarray}
    S=-k_B \Tr \left[ \hat{\rho} \ln(\hat{\rho}) \right],
\end{eqnarray}
where $\hat{\rho}$ is the reduced density operator. For our case, using the last equation it is easy to show that
\begin{eqnarray}\label{Entropy_2}
    S=k_B \ln(2).
\end{eqnarray}
It is quite remarkable to see from Eq.~\eqref{Entropy_1} that the quantum corrections to the Hawking-Bekenstein entropy are of logarithmic form just like the entanglement entropy. By equating Eqs.~\eqref{Entropy_1} and \eqref{Entropy_2}, we get the minimal length 
\begin{equation}
    l_0 = \pm \frac{\sqrt{3}}{3} \left(\frac{\ln(2)}{ \pi \ln(\sqrt{2}+\sqrt{3}) }\right)^{1/2}  \sqrt{\frac{G \hbar}{c^3}}
\end{equation}
and thus we found another argument given by the last equation confirming the fact that $l_0$ is of the Planck length order. More precisely, we get 
\begin{equation}\label{zero-point_length}
  l_0 = \pm \frac{\sqrt{3}}{3} \left(\frac{\ln(2)}{ \pi \ln(\sqrt{2}+\sqrt{3}) }\right)^{1/2}  l_{Pl}=0.253 \, l_{Pl}
\end{equation}
where $l_{Pl}=\sqrt{\frac{G \hbar}{c^3}}=1.61 \times 10^{-35}$. For the $e^+ e^-$ pair we neglect the electric charge as must be zero, so we consider only the uncharged ER bridge. The horizon and the mass of such a particle-black hole are found by combing  $r=\sqrt{2} l_0$ and $M=3 \sqrt{3} l_0/4$ [restoring the constants $G$ and $c$], yielding
\begin{eqnarray}
    r=\frac{4\, \sqrt{2}}{3 \sqrt{3}}\frac{G M}{c^2},
\end{eqnarray}
from where do we get the mass
\begin{eqnarray}
    M =\frac{ 3\, \sqrt{3}\,l_0\, c^2}{4 \,\,G}.
\end{eqnarray}
The last equation for the mass is therefore proportional to the zero-point length. We can rewrite as 
\begin{eqnarray}
    M_{\pm}= \pm \frac{3\,c^2}{4\,G} \left(\frac{\ln(2)}{ \pi \ln(\sqrt{2}+\sqrt{3}) }\right)^{1/2}   \sqrt{\frac{G \hbar}{c^3}}.
\end{eqnarray}
This relation can be expressed also in terms of the Planck mass 
\begin{eqnarray}
     M_{\pm}= \pm \frac{3}{4}  \left(\frac{\ln(2)}{ \pi \ln(\sqrt{2}+\sqrt{3}) }\right)^{1/2}  M_{Pl}
\end{eqnarray}
where $M_{Pl}=\sqrt{\frac{\hbar c}{G}}$ is the Planck mass. Taking only the positive solution for the mass, we get 
\begin{eqnarray}
     M_{\pm}= \pm  7.24 \times 10^{-9} \rm kg,
\end{eqnarray}
which is in agreement with the result in \cite{dai2020testing}. It is quite interesting to observe that this bound for the mass coincides with the mass bound that was found from the Casimir energy in Eq.~\eqref{WH_mass}.  The fact that we get a mass close to the Planck  mass may be related to the fact that gravity is self-complete, namely, it protects the ultraviolet regime
by setting quantum mechanical limits to length and energy/mass \cite{Isi:2013cxa}. This means that $10^{-9}$~kg is much larger than a typical elementary particle pair’s mass and rises the problem that such a wormhole can be created as a consequence of entanglement. However, for particle collisions, if a particle is accelerated to an energy larger than extreme mass, it is compressed below its corresponding horizon and the particle can undergo a gravitational collapse into a black hole. This may imply that ER=EPR can only exist in special configurations, namely if we take a pair of such particles that undergoes a gravitational collapse into a black hole, as for primordial black holes \cite{carr1974black,hawking1971gravitationally}, and only such entangled configuration results in ER=EPR. Moreover, the negative energy of such wormholes is directly provided by the two negative energy solutions of the Dirac equation for primordial antimatter, see for instance \cite{debergh2018evidence}.

Let us now elaborate the case of wormhole spacetimes without a horizon connecting two particles. The mass value $10^{-9}$~kg is closely linked to the presence of the horizon. In principle, $M$ is a free parameter, hence we can avoid the presence of horizon as we explained in Section~\ref{Sec_ERbridge}, simply by imposing the condition for the mass parameter
\begin{equation}
    M < \left|\pm \frac{3 \sqrt{3} l_0 c^2}{4 G}\right|.
\end{equation}
There is another constrain for the region of such a regular spacetime; namely, we have to exclude the point $r=0$ or $u=\pm l_0$ [see Eq.~\eqref{throat}] to avoid the singularity at this point. A physical acceptable region that we can choose is $r\geq l_0$, thus in terms of $u$ the region should be $u\in \left(-\infty, - \sqrt{2}l_0\right] \cup \left[ \sqrt{2}l_0, \infty\right)$.  In the case of horizonless spacetime, the Bekenstein-Hawking entropy is not a well-defined quantity, but we can still use the entanglement entropy given by Eq.~\eqref{Entropy_1} to compute the entropy of the entangled pair. By choosing an arbitrarily small mass, say for example a mass compatible with the electron rest mass $M \sim 10^{-30}$~kg, we can avoid the extreme mass value problem found above, for wormholes with horizonless spacetime and we have the upper mass bound $M < 10^{-9}$~kg. If such wormholes exist during vacuum fluctuations, they exist for a short amount of time in accordance with the energy-time uncertainty principle.

\begin{figure}
    \centering
    \includegraphics[scale=0.55]{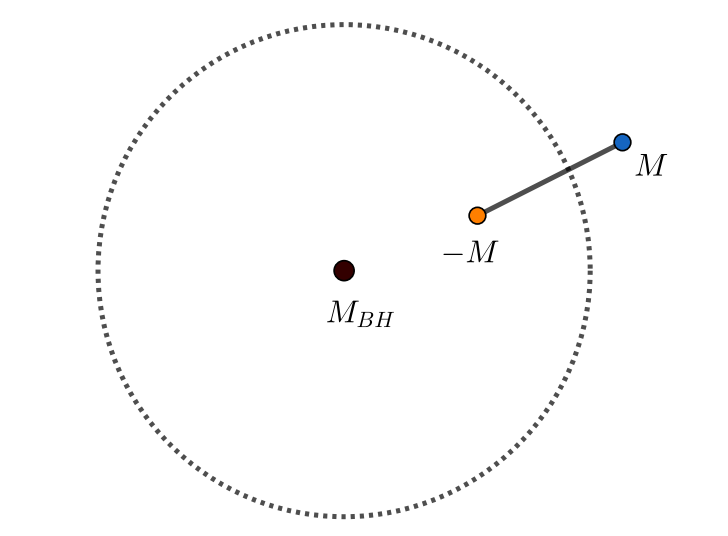}
   \caption{Pair of entangled particles, one inside and the other outside the horizon.}
   \label{fig:radiation}
\end{figure}

A peculiar case of quantum fluctuations however occurs near a black hole where the relation $\Delta E \,\Delta t \sim \hbar$ still holds. Here, instead of total energy fluctuation of the pair $\Delta E \sim 2 M  c^2$ where $M$ is the rest mass of the particle, one can have $\Delta E\to 0$ \cite{Mathur:2008wi}. The net energy of the particle inside the event horizon can be negative and the total energy of the entangled system is zero. In particular, for the particle that tunnels inside the black hole we have the net energy 
\begin{eqnarray}
    E=M c^2-\frac{G M_{BH} M}{\sqrt{r^2+l_0^2}}+\rm KE. \label{energy}
\end{eqnarray}
For simplicity, we take the kinetic energy to be zero, i.e. $\rm KE=0$, and we have modified the potential energy due to the zero-point length. The net energy is, therefore, zero if 
\begin{eqnarray}
    r_0= \frac{\sqrt{G^2M_{BH}^2-l_0^2 c^4}}{c^2}\simeq \frac{G M_{BH}}{c^2}
\end{eqnarray}
since $l_0<<M_{BH}$. This simple calculation shows that for $r<r_0$, the energy of the particle is indeed negative, i.e. $E<0$, and we can see this from Eq.~\eqref{energy}. By taking both solutions the total energy/mass is zero
\begin{eqnarray}
     E_{\text{total}}=M_{+}c^2+M_{-}c^2 \simeq 0.
\end{eqnarray}
This means we can avoid the extreme mass value problem which we faced in this section. Note also that, quantum mechanically, the total energy does not mean to be precisely zero, but it can also be close to zero, say of the order of a typical elementary particle pair’s mass, i.e. $MeV$. In this way, the interior of the black hole with total area $A \sim N 8 \pi l_0^2$, and total entropy $A \sim N k_B \ln(2)$ where $N$ is the number of total quanta emitted by the black hole, becomes entangled with the exterior region via Hawking radiation, see Fig~\ref{fig:radiation}.
This may explain why the universe is composed of matter rather than antimatter. This also resonates with the timeless state of the universe \cite{Ali:2021ela}. In fact, it was shown in \cite{Jusufi:2023pzt} how the final state of a gravitational collapse under zero-point length (the collapse of matter) leads to a black hole with a very dense inner core without a central singularity.  One of the possibilities for the final state after the total evaporation of the black hole is the remnant state (particle) which has exact mass $M$.  This is very interesting as it offers a new way to tackle the black hole information paradox with entangled wormholes, as proposed in \cite{Almheiri:2019qdq}.

\section{Conclusion}\label{Sec_Conclusion}

We used a string T-duality corrected pair of regular black holes and obtained an ER bridge involving negative energy/mass with the wormhole throat proportional to the Planck-scale zero-point length. We point out that for an extreme mass configuration, a pair of particle-black hole configurations, we can have an ER bridge with an area horizon that coincides with the Bekenstein minimal area bound along with a wormhole mass proportional to the Planck mass. This could be related to the gravitational self-completeness feature with quantum mechanical limits to the mass. We argue that Hawking radiation could be the best way to the geometric realization of quantum entanglement for particle/antiparticle pairs emitted by black holes. We also discuss the ER bridge for horizonless wormholes in the mass-compatible particle sector. In such a case, the horizon is avoided by choosing an arbitrarily small mass that could be generated in principle by quantum fluctuations. We closely inspect how such a negative mass can arise from the Casimir effect. Finally, this can shed new light on the ER=EPR conjecture in a $(3+1)-$dimensional theory of gravity.

\bibliographystyle{apsrev4-1}
\bibliography{wormhole}

\begin{thebibliography}{53}%
\makeatletter
\providecommand \@ifxundefined [1]{%
 \@ifx{#1\undefined}
}%
\providecommand \@ifnum [1]{%
 \ifnum #1\expandafter \@firstoftwo
 \else \expandafter \@secondoftwo
 \fi
}%
\providecommand \@ifx [1]{%
 \ifx #1\expandafter \@firstoftwo
 \else \expandafter \@secondoftwo
 \fi
}%
\providecommand \natexlab [1]{#1}%
\providecommand \enquote  [1]{``#1''}%
\providecommand \bibnamefont  [1]{#1}%
\providecommand \bibfnamefont [1]{#1}%
\providecommand \citenamefont [1]{#1}%
\providecommand \href@noop [0]{\@secondoftwo}%
\providecommand \href [0]{\begingroup \@sanitize@url \@href}%
\providecommand \@href[1]{\@@startlink{#1}\@@href}%
\providecommand \@@href[1]{\endgroup#1\@@endlink}%
\providecommand \@sanitize@url [0]{\catcode `\\12\catcode `\$12\catcode
  `\&12\catcode `\#12\catcode `\^12\catcode `\_12\catcode `\%12\relax}%
\providecommand \@@startlink[1]{}%
\providecommand \@@endlink[0]{}%
\providecommand \url  [0]{\begingroup\@sanitize@url \@url }%
\providecommand \@url [1]{\endgroup\@href {#1}{\urlprefix }}%
\providecommand \urlprefix  [0]{URL }%
\providecommand \Eprint [0]{\href }%
\providecommand \doibase [0]{http://dx.doi.org/}%
\providecommand \selectlanguage [0]{\@gobble}%
\providecommand \bibinfo  [0]{\@secondoftwo}%
\providecommand \bibfield  [0]{\@secondoftwo}%
\providecommand \translation [1]{[#1]}%
\providecommand \BibitemOpen [0]{}%
\providecommand \bibitemStop [0]{}%
\providecommand \bibitemNoStop [0]{.\EOS\space}%
\providecommand \EOS [0]{\spacefactor3000\relax}%
\providecommand \BibitemShut  [1]{\csname bibitem#1\endcsname}%
\let\auto@bib@innerbib\@empty
\bibitem [{\citenamefont {Maldacena}\ and\ \citenamefont
  {Susskind}(2013)}]{maldacena2013cool}%
  \BibitemOpen
  \bibfield  {author} {\bibinfo {author} {\bibfnamefont {J.}~\bibnamefont
  {Maldacena}}\ and\ \bibinfo {author} {\bibfnamefont {L.}~\bibnamefont
  {Susskind}},\ }\href {\doibase 10.1002/prop.201300020} {\bibfield  {journal}
  {\bibinfo  {journal} {Fortsch. Phys.}\ }\textbf {\bibinfo {volume} {61}},\
  \bibinfo {pages} {781} (\bibinfo {year} {2013})},\ \Eprint
  {http://arxiv.org/abs/1306.0533} {arXiv:1306.0533 [hep-th]} \BibitemShut
  {NoStop}%
\bibitem [{\citenamefont {Einstein}\ and\ \citenamefont
  {Rosen}(1935)}]{einstein1935particle}%
  \BibitemOpen
  \bibfield  {author} {\bibinfo {author} {\bibfnamefont {A.}~\bibnamefont
  {Einstein}}\ and\ \bibinfo {author} {\bibfnamefont {N.}~\bibnamefont
  {Rosen}},\ }\href {\doibase 10.1103/PhysRev.48.73} {\bibfield  {journal}
  {\bibinfo  {journal} {Phys. Rev.}\ }\textbf {\bibinfo {volume} {48}},\
  \bibinfo {pages} {73} (\bibinfo {year} {1935})}\BibitemShut {NoStop}%
\bibitem [{\citenamefont {Bengtsson}\ and\ \citenamefont
  {{\.Z}yczkowski}(2017)}]{bengtsson2017geometry}%
  \BibitemOpen
  \bibfield  {author} {\bibinfo {author} {\bibfnamefont {I.}~\bibnamefont
  {Bengtsson}}\ and\ \bibinfo {author} {\bibfnamefont {K.}~\bibnamefont
  {{\.Z}yczkowski}},\ }\href@noop {} {\emph {\bibinfo {title} {Geometry of
  quantum states: an introduction to quantum entanglement}}}\ (\bibinfo
  {publisher} {Cambridge University Press},\ \bibinfo {year}
  {2017})\BibitemShut {NoStop}%
\bibitem [{\citenamefont {Einstein}\ \emph {et~al.}(1935)\citenamefont
  {Einstein}, \citenamefont {Podolsky},\ and\ \citenamefont
  {Rosen}}]{einstein1935can}%
  \BibitemOpen
  \bibfield  {author} {\bibinfo {author} {\bibfnamefont {A.}~\bibnamefont
  {Einstein}}, \bibinfo {author} {\bibfnamefont {B.}~\bibnamefont {Podolsky}},
  \ and\ \bibinfo {author} {\bibfnamefont {N.}~\bibnamefont {Rosen}},\ }\href
  {\doibase 10.1103/PhysRev.47.777} {\bibfield  {journal} {\bibinfo  {journal}
  {Phys. Rev.}\ }\textbf {\bibinfo {volume} {47}},\ \bibinfo {pages} {777}
  (\bibinfo {year} {1935})}\BibitemShut {NoStop}%
\bibitem [{\citenamefont {Gao}\ and\ \citenamefont
  {Jafferis}(2021)}]{gao2021traversable}%
  \BibitemOpen
  \bibfield  {author} {\bibinfo {author} {\bibfnamefont {P.}~\bibnamefont
  {Gao}}\ and\ \bibinfo {author} {\bibfnamefont {D.~L.}\ \bibnamefont
  {Jafferis}},\ }\href@noop {} {\bibfield  {journal} {\bibinfo  {journal}
  {Journal of High Energy Physics}\ }\textbf {\bibinfo {volume} {2021}},\
  \bibinfo {pages} {1} (\bibinfo {year} {2021})}\BibitemShut {NoStop}%
\bibitem [{\citenamefont {Maldacena}\ and\ \citenamefont
  {Stanford}(2016)}]{maldacena2016remarks}%
  \BibitemOpen
  \bibfield  {author} {\bibinfo {author} {\bibfnamefont {J.}~\bibnamefont
  {Maldacena}}\ and\ \bibinfo {author} {\bibfnamefont {D.}~\bibnamefont
  {Stanford}},\ }\href@noop {} {\bibfield  {journal} {\bibinfo  {journal}
  {Physical Review D}\ }\textbf {\bibinfo {volume} {94}},\ \bibinfo {pages}
  {106002} (\bibinfo {year} {2016})}\BibitemShut {NoStop}%
\bibitem [{\citenamefont {Maldacena}\ and\ \citenamefont
  {Milekhin}(2021)}]{maldacena2021syk}%
  \BibitemOpen
  \bibfield  {author} {\bibinfo {author} {\bibfnamefont {J.}~\bibnamefont
  {Maldacena}}\ and\ \bibinfo {author} {\bibfnamefont {A.}~\bibnamefont
  {Milekhin}},\ }\href@noop {} {\bibfield  {journal} {\bibinfo  {journal}
  {Journal of High Energy Physics}\ }\textbf {\bibinfo {volume} {2021}},\
  \bibinfo {pages} {1} (\bibinfo {year} {2021})}\BibitemShut {NoStop}%
\bibitem [{\citenamefont {Sachdev}\ and\ \citenamefont
  {Ye}(1993)}]{Sachdev:1992fk}%
  \BibitemOpen
  \bibfield  {author} {\bibinfo {author} {\bibfnamefont {S.}~\bibnamefont
  {Sachdev}}\ and\ \bibinfo {author} {\bibfnamefont {J.}~\bibnamefont {Ye}},\
  }\href {\doibase 10.1103/PhysRevLett.70.3339} {\bibfield  {journal} {\bibinfo
   {journal} {Phys. Rev. Lett.}\ }\textbf {\bibinfo {volume} {70}},\ \bibinfo
  {pages} {3339} (\bibinfo {year} {1993})},\ \Eprint
  {http://arxiv.org/abs/cond-mat/9212030} {arXiv:cond-mat/9212030} \BibitemShut
  {NoStop}%
\bibitem [{\citenamefont {Kitaev}(2015)}]{kitaev2015simple}%
  \BibitemOpen
  \bibfield  {author} {\bibinfo {author} {\bibfnamefont {A.}~\bibnamefont
  {Kitaev}},\ }\href@noop {} {\bibfield  {journal} {\bibinfo  {journal}
  {Entanglement in Strongly-Correlated Quantum Matter}\ ,\ \bibinfo {pages}
  {38}} (\bibinfo {year} {2015})}\BibitemShut {NoStop}%
\bibitem [{\citenamefont {Jafferis}\ \emph
  {et~al.}(2022{\natexlab{a}})\citenamefont {Jafferis}, \citenamefont
  {Zlokapa}, \citenamefont {Lykken}, \citenamefont {Kolchmeyer}, \citenamefont
  {Davis}, \citenamefont {Lauk}, \citenamefont {Neven},\ and\ \citenamefont
  {Spiropulu}}]{Jafferis2022}%
  \BibitemOpen
  \bibfield  {author} {\bibinfo {author} {\bibfnamefont {D.}~\bibnamefont
  {Jafferis}}, \bibinfo {author} {\bibfnamefont {A.}~\bibnamefont {Zlokapa}},
  \bibinfo {author} {\bibfnamefont {J.~D.}\ \bibnamefont {Lykken}}, \bibinfo
  {author} {\bibfnamefont {D.~K.}\ \bibnamefont {Kolchmeyer}}, \bibinfo
  {author} {\bibfnamefont {S.~I.}\ \bibnamefont {Davis}}, \bibinfo {author}
  {\bibfnamefont {N.}~\bibnamefont {Lauk}}, \bibinfo {author} {\bibfnamefont
  {H.}~\bibnamefont {Neven}}, \ and\ \bibinfo {author} {\bibfnamefont
  {M.}~\bibnamefont {Spiropulu}},\ }\href {\doibase 10.1038/s41586-022-05424-3}
  {\bibfield  {journal} {\bibinfo  {journal} {Nature}\ }\textbf {\bibinfo
  {volume} {612}},\ \bibinfo {pages} {51} (\bibinfo {year}
  {2022}{\natexlab{a}})}\BibitemShut {NoStop}%
\bibitem [{\citenamefont {Dai}\ \emph {et~al.}(2020{\natexlab{a}})\citenamefont
  {Dai}, \citenamefont {Minic}, \citenamefont {Stojkovic},\ and\ \citenamefont
  {Fu}}]{dai2020testing}%
  \BibitemOpen
  \bibfield  {author} {\bibinfo {author} {\bibfnamefont {D.-C.}\ \bibnamefont
  {Dai}}, \bibinfo {author} {\bibfnamefont {D.}~\bibnamefont {Minic}}, \bibinfo
  {author} {\bibfnamefont {D.}~\bibnamefont {Stojkovic}}, \ and\ \bibinfo
  {author} {\bibfnamefont {C.}~\bibnamefont {Fu}},\ }\href@noop {} {\bibfield
  {journal} {\bibinfo  {journal} {Physical Review D}\ }\textbf {\bibinfo
  {volume} {102}},\ \bibinfo {pages} {066004} (\bibinfo {year}
  {2020}{\natexlab{a}})}\BibitemShut {NoStop}%
\bibitem [{\citenamefont {Ali}\ \emph {et~al.}(2022{\natexlab{a}})\citenamefont
  {Ali}, \citenamefont {Moulay}, \citenamefont {Jusufi},\ and\ \citenamefont
  {Alshal}}]{ali2022unitary}%
  \BibitemOpen
  \bibfield  {author} {\bibinfo {author} {\bibfnamefont {A.~F.}\ \bibnamefont
  {Ali}}, \bibinfo {author} {\bibfnamefont {E.}~\bibnamefont {Moulay}},
  \bibinfo {author} {\bibfnamefont {K.}~\bibnamefont {Jusufi}}, \ and\ \bibinfo
  {author} {\bibfnamefont {H.}~\bibnamefont {Alshal}},\ }\href {\doibase
  10.1140/epjc/s10052-022-11095-1} {\bibfield  {journal} {\bibinfo  {journal}
  {The European Physical Journal C}\ }\textbf {\bibinfo {volume} {82}},\
  \bibinfo {pages} {1170} (\bibinfo {year} {2022}{\natexlab{a}})}\BibitemShut
  {NoStop}%
\bibitem [{\citenamefont {Bertlmann}(2017)}]{bertlmann2017bell}%
  \BibitemOpen
  \bibfield  {author} {\bibinfo {author} {\bibfnamefont {R.}~\bibnamefont
  {Bertlmann}},\ }in\ \href@noop {} {\emph {\bibinfo {booktitle} {Quantum [Un]
  Speakables II}}}\ (\bibinfo  {publisher} {Springer},\ \bibinfo {year}
  {2017})\ pp.\ \bibinfo {pages} {17--80}\BibitemShut {NoStop}%
\bibitem [{\citenamefont {Davies}\ \emph {et~al.}(1976)\citenamefont {Davies},
  \citenamefont {Fulling},\ and\ \citenamefont {Unruh}}]{davies1976energy}%
  \BibitemOpen
  \bibfield  {author} {\bibinfo {author} {\bibfnamefont {P.~C.}\ \bibnamefont
  {Davies}}, \bibinfo {author} {\bibfnamefont {S.~A.}\ \bibnamefont {Fulling}},
  \ and\ \bibinfo {author} {\bibfnamefont {W.~G.}\ \bibnamefont {Unruh}},\
  }\href@noop {} {\bibfield  {journal} {\bibinfo  {journal} {Physical Review
  D}\ }\textbf {\bibinfo {volume} {13}},\ \bibinfo {pages} {2720} (\bibinfo
  {year} {1976})}\BibitemShut {NoStop}%
\bibitem [{\citenamefont {Hawking}(1975)}]{hawking1975particle}%
  \BibitemOpen
  \bibfield  {author} {\bibinfo {author} {\bibfnamefont {S.~W.}\ \bibnamefont
  {Hawking}},\ }\href@noop {} {\bibfield  {journal} {\bibinfo  {journal}
  {Communications in Mathematical Physics}\ }\textbf {\bibinfo {volume} {43}},\
  \bibinfo {pages} {199} (\bibinfo {year} {1975})}\BibitemShut {NoStop}%
\bibitem [{\citenamefont {Summers}(2011)}]{summers2011yet}%
  \BibitemOpen
  \bibfield  {author} {\bibinfo {author} {\bibfnamefont {S.~J.}\ \bibnamefont
  {Summers}},\ }in\ \href@noop {} {\emph {\bibinfo {booktitle} {Deep Beauty}}}\
  (\bibinfo  {publisher} {Cambridge University Press},\ \bibinfo {year}
  {2011})\ pp.\ \bibinfo {pages} {317--341}\BibitemShut {NoStop}%
\bibitem [{\citenamefont {Dalvit}\ \emph {et~al.}(2011)\citenamefont {Dalvit},
  \citenamefont {Milonni}, \citenamefont {Roberts},\ and\ \citenamefont
  {Da~Rosa}}]{dalvit2011casimir}%
  \BibitemOpen
  \bibfield  {author} {\bibinfo {author} {\bibfnamefont {D.}~\bibnamefont
  {Dalvit}}, \bibinfo {author} {\bibfnamefont {P.}~\bibnamefont {Milonni}},
  \bibinfo {author} {\bibfnamefont {D.}~\bibnamefont {Roberts}}, \ and\
  \bibinfo {author} {\bibfnamefont {F.}~\bibnamefont {Da~Rosa}},\ }\href@noop
  {} {\emph {\bibinfo {title} {Casimir physics}}},\ \bibinfo {series} {Lecture
  Notes in Physics}, Vol.\ \bibinfo {volume} {834}\ (\bibinfo  {publisher}
  {Springer},\ \bibinfo {year} {2011})\BibitemShut {NoStop}%
\bibitem [{\citenamefont {Nicolini}\ \emph {et~al.}(2019)\citenamefont
  {Nicolini}, \citenamefont {Spallucci},\ and\ \citenamefont
  {Wondrak}}]{Nicolini:2019irw}%
  \BibitemOpen
  \bibfield  {author} {\bibinfo {author} {\bibfnamefont {P.}~\bibnamefont
  {Nicolini}}, \bibinfo {author} {\bibfnamefont {E.}~\bibnamefont {Spallucci}},
  \ and\ \bibinfo {author} {\bibfnamefont {M.~F.}\ \bibnamefont {Wondrak}},\
  }\href {\doibase 10.1016/j.physletb.2019.134888} {\bibfield  {journal}
  {\bibinfo  {journal} {Phys. Lett. B}\ }\textbf {\bibinfo {volume} {797}},\
  \bibinfo {pages} {134888} (\bibinfo {year} {2019})},\ \Eprint
  {http://arxiv.org/abs/1902.11242} {arXiv:1902.11242 [gr-qc]} \BibitemShut
  {NoStop}%
\bibitem [{\citenamefont {Simpson}\ and\ \citenamefont
  {Visser}(2019)}]{simpson2019black}%
  \BibitemOpen
  \bibfield  {author} {\bibinfo {author} {\bibfnamefont {A.}~\bibnamefont
  {Simpson}}\ and\ \bibinfo {author} {\bibfnamefont {M.}~\bibnamefont
  {Visser}},\ }\href@noop {} {\bibfield  {journal} {\bibinfo  {journal}
  {Journal of Cosmology and Astroparticle Physics}\ }\textbf {\bibinfo {volume}
  {2019}},\ \bibinfo {pages} {042} (\bibinfo {year} {2019})}\BibitemShut
  {NoStop}%
\bibitem [{\citenamefont {Morris}\ and\ \citenamefont
  {Thorne}(1988)}]{morris1988wormholes2}%
  \BibitemOpen
  \bibfield  {author} {\bibinfo {author} {\bibfnamefont {M.~S.}\ \bibnamefont
  {Morris}}\ and\ \bibinfo {author} {\bibfnamefont {K.~S.}\ \bibnamefont
  {Thorne}},\ }\href@noop {} {\bibfield  {journal} {\bibinfo  {journal}
  {American Journal of Physics}\ }\textbf {\bibinfo {volume} {56}},\ \bibinfo
  {pages} {395} (\bibinfo {year} {1988})}\BibitemShut {NoStop}%
\bibitem [{\citenamefont {Ali}\ \emph {et~al.}(2009)\citenamefont {Ali},
  \citenamefont {Das},\ and\ \citenamefont {Vagenas}}]{Ali:2009zq}%
  \BibitemOpen
  \bibfield  {author} {\bibinfo {author} {\bibfnamefont {A.~F.}\ \bibnamefont
  {Ali}}, \bibinfo {author} {\bibfnamefont {S.}~\bibnamefont {Das}}, \ and\
  \bibinfo {author} {\bibfnamefont {E.~C.}\ \bibnamefont {Vagenas}},\ }\href
  {\doibase 10.1016/j.physletb.2009.06.061} {\bibfield  {journal} {\bibinfo
  {journal} {Phys. Lett. B}\ }\textbf {\bibinfo {volume} {678}},\ \bibinfo
  {pages} {497} (\bibinfo {year} {2009})},\ \Eprint
  {http://arxiv.org/abs/0906.5396} {arXiv:0906.5396 [hep-th]} \BibitemShut
  {NoStop}%
\bibitem [{\citenamefont {Garay}(1995)}]{Garay:1994en}%
  \BibitemOpen
  \bibfield  {author} {\bibinfo {author} {\bibfnamefont {L.~J.}\ \bibnamefont
  {Garay}},\ }\href {\doibase 10.1142/S0217751X95000085} {\bibfield  {journal}
  {\bibinfo  {journal} {Int. J. Mod. Phys. A}\ }\textbf {\bibinfo {volume}
  {10}},\ \bibinfo {pages} {145} (\bibinfo {year} {1995})},\ \Eprint
  {http://arxiv.org/abs/gr-qc/9403008} {arXiv:gr-qc/9403008} \BibitemShut
  {NoStop}%
\bibitem [{\citenamefont {Amati}\ \emph {et~al.}(1989)\citenamefont {Amati},
  \citenamefont {Ciafaloni},\ and\ \citenamefont {Veneziano}}]{Amati:1988tn}%
  \BibitemOpen
  \bibfield  {author} {\bibinfo {author} {\bibfnamefont {D.}~\bibnamefont
  {Amati}}, \bibinfo {author} {\bibfnamefont {M.}~\bibnamefont {Ciafaloni}}, \
  and\ \bibinfo {author} {\bibfnamefont {G.}~\bibnamefont {Veneziano}},\ }\href
  {\doibase 10.1016/0370-2693(89)91366-X} {\bibfield  {journal} {\bibinfo
  {journal} {Phys. Lett. B}\ }\textbf {\bibinfo {volume} {216}},\ \bibinfo
  {pages} {41} (\bibinfo {year} {1989})}\BibitemShut {NoStop}%
\bibitem [{\citenamefont {Hochberg}\ and\ \citenamefont
  {Visser}(1997)}]{hochberg1997geometric}%
  \BibitemOpen
  \bibfield  {author} {\bibinfo {author} {\bibfnamefont {D.}~\bibnamefont
  {Hochberg}}\ and\ \bibinfo {author} {\bibfnamefont {M.}~\bibnamefont
  {Visser}},\ }\href@noop {} {\bibfield  {journal} {\bibinfo  {journal}
  {Physical Review D}\ }\textbf {\bibinfo {volume} {56}},\ \bibinfo {pages}
  {4745} (\bibinfo {year} {1997})}\BibitemShut {NoStop}%
\bibitem [{\citenamefont {Hochberg}\ and\ \citenamefont
  {Visser}(1998)}]{hochberg1998dynamic}%
  \BibitemOpen
  \bibfield  {author} {\bibinfo {author} {\bibfnamefont {D.}~\bibnamefont
  {Hochberg}}\ and\ \bibinfo {author} {\bibfnamefont {M.}~\bibnamefont
  {Visser}},\ }\href@noop {} {\bibfield  {journal} {\bibinfo  {journal}
  {Physical Review D}\ }\textbf {\bibinfo {volume} {58}},\ \bibinfo {pages}
  {044021} (\bibinfo {year} {1998})}\BibitemShut {NoStop}%
\bibitem [{\citenamefont {Ida}\ and\ \citenamefont
  {Hayward}(1999)}]{ida1999much}%
  \BibitemOpen
  \bibfield  {author} {\bibinfo {author} {\bibfnamefont {D.}~\bibnamefont
  {Ida}}\ and\ \bibinfo {author} {\bibfnamefont {S.~A.}\ \bibnamefont
  {Hayward}},\ }\href@noop {} {\bibfield  {journal} {\bibinfo  {journal}
  {Physics Letters A}\ }\textbf {\bibinfo {volume} {260}},\ \bibinfo {pages}
  {175} (\bibinfo {year} {1999})}\BibitemShut {NoStop}%
\bibitem [{\citenamefont {Morris}\ \emph {et~al.}(1988)\citenamefont {Morris},
  \citenamefont {Thorne},\ and\ \citenamefont
  {Yurtsever}}]{morris1988wormholes}%
  \BibitemOpen
  \bibfield  {author} {\bibinfo {author} {\bibfnamefont {M.~S.}\ \bibnamefont
  {Morris}}, \bibinfo {author} {\bibfnamefont {K.~S.}\ \bibnamefont {Thorne}},
  \ and\ \bibinfo {author} {\bibfnamefont {U.}~\bibnamefont {Yurtsever}},\
  }\href@noop {} {\bibfield  {journal} {\bibinfo  {journal} {Physical Review
  Letters}\ }\textbf {\bibinfo {volume} {61}},\ \bibinfo {pages} {1446}
  (\bibinfo {year} {1988})}\BibitemShut {NoStop}%
\bibitem [{\citenamefont {Bambi}\ and\ \citenamefont
  {Stojkovic}(2021)}]{Bambi:2021qfo}%
  \BibitemOpen
  \bibfield  {author} {\bibinfo {author} {\bibfnamefont {C.}~\bibnamefont
  {Bambi}}\ and\ \bibinfo {author} {\bibfnamefont {D.}~\bibnamefont
  {Stojkovic}},\ }\href {\doibase 10.3390/universe7050136} {\bibfield
  {journal} {\bibinfo  {journal} {Universe}\ }\textbf {\bibinfo {volume} {7}},\
  \bibinfo {pages} {136} (\bibinfo {year} {2021})},\ \Eprint
  {http://arxiv.org/abs/2105.00881} {arXiv:2105.00881 [gr-qc]} \BibitemShut
  {NoStop}%
\bibitem [{\citenamefont {Dai}\ \emph {et~al.}(2020{\natexlab{b}})\citenamefont
  {Dai}, \citenamefont {Minic},\ and\ \citenamefont {Stojkovic}}]{Dai:2020rnc}%
  \BibitemOpen
  \bibfield  {author} {\bibinfo {author} {\bibfnamefont {D.-C.}\ \bibnamefont
  {Dai}}, \bibinfo {author} {\bibfnamefont {D.}~\bibnamefont {Minic}}, \ and\
  \bibinfo {author} {\bibfnamefont {D.}~\bibnamefont {Stojkovic}},\ }\href
  {\doibase 10.1140/epjc/s10052-020-08698-x} {\bibfield  {journal} {\bibinfo
  {journal} {Eur. Phys. J. C}\ }\textbf {\bibinfo {volume} {80}},\ \bibinfo
  {pages} {1103} (\bibinfo {year} {2020}{\natexlab{b}})},\ \Eprint
  {http://arxiv.org/abs/2010.03947} {arXiv:2010.03947 [gr-qc]} \BibitemShut
  {NoStop}%
\bibitem [{\citenamefont {Simonetti}\ \emph {et~al.}(2021)\citenamefont
  {Simonetti}, \citenamefont {Kavic}, \citenamefont {Minic}, \citenamefont
  {Stojkovic},\ and\ \citenamefont {Dai}}]{Simonetti:2020ivl}%
  \BibitemOpen
  \bibfield  {author} {\bibinfo {author} {\bibfnamefont {J.~H.}\ \bibnamefont
  {Simonetti}}, \bibinfo {author} {\bibfnamefont {M.~J.}\ \bibnamefont
  {Kavic}}, \bibinfo {author} {\bibfnamefont {D.}~\bibnamefont {Minic}},
  \bibinfo {author} {\bibfnamefont {D.}~\bibnamefont {Stojkovic}}, \ and\
  \bibinfo {author} {\bibfnamefont {D.-C.}\ \bibnamefont {Dai}},\ }\href
  {\doibase 10.1103/PhysRevD.104.L081502} {\bibfield  {journal} {\bibinfo
  {journal} {Phys. Rev. D}\ }\textbf {\bibinfo {volume} {104}},\ \bibinfo
  {pages} {L081502} (\bibinfo {year} {2021})},\ \Eprint
  {http://arxiv.org/abs/2007.12184} {arXiv:2007.12184 [gr-qc]} \BibitemShut
  {NoStop}%
\bibitem [{\citenamefont {Jusufi}\ \emph {et~al.}(2020)\citenamefont {Jusufi},
  \citenamefont {Banerjee},\ and\ \citenamefont {Ghosh}}]{Jusufi:2020yus}%
  \BibitemOpen
  \bibfield  {author} {\bibinfo {author} {\bibfnamefont {K.}~\bibnamefont
  {Jusufi}}, \bibinfo {author} {\bibfnamefont {A.}~\bibnamefont {Banerjee}}, \
  and\ \bibinfo {author} {\bibfnamefont {S.~G.}\ \bibnamefont {Ghosh}},\ }\href
  {\doibase 10.1140/epjc/s10052-020-8287-x} {\bibfield  {journal} {\bibinfo
  {journal} {Eur. Phys. J. C}\ }\textbf {\bibinfo {volume} {80}},\ \bibinfo
  {pages} {698} (\bibinfo {year} {2020})},\ \Eprint
  {http://arxiv.org/abs/2004.10750} {arXiv:2004.10750 [gr-qc]} \BibitemShut
  {NoStop}%
\bibitem [{\citenamefont {Jusufi}(2018)}]{Jusufi:2018waj}%
  \BibitemOpen
  \bibfield  {author} {\bibinfo {author} {\bibfnamefont {K.}~\bibnamefont
  {Jusufi}},\ }\href {\doibase 10.1103/PhysRevD.98.044016} {\bibfield
  {journal} {\bibinfo  {journal} {Phys. Rev. D}\ }\textbf {\bibinfo {volume}
  {98}},\ \bibinfo {pages} {044016} (\bibinfo {year} {2018})},\ \Eprint
  {http://arxiv.org/abs/1803.02317} {arXiv:1803.02317 [gr-qc]} \BibitemShut
  {NoStop}%
\bibitem [{\citenamefont {Jusufi}\ \emph {et~al.}(2022)\citenamefont {Jusufi},
  \citenamefont {Kumar}, \citenamefont {Azreg-A\"\i{}nou}, \citenamefont
  {Jamil}, \citenamefont {Wu},\ and\ \citenamefont {Bambi}}]{Jusufi:2021lei}%
  \BibitemOpen
  \bibfield  {author} {\bibinfo {author} {\bibfnamefont {K.}~\bibnamefont
  {Jusufi}}, \bibinfo {author} {\bibfnamefont {S.}~\bibnamefont {Kumar}},
  \bibinfo {author} {\bibfnamefont {M.}~\bibnamefont {Azreg-A\"\i{}nou}},
  \bibinfo {author} {\bibfnamefont {M.}~\bibnamefont {Jamil}}, \bibinfo
  {author} {\bibfnamefont {Q.}~\bibnamefont {Wu}}, \ and\ \bibinfo {author}
  {\bibfnamefont {C.}~\bibnamefont {Bambi}},\ }\href {\doibase
  10.1140/epjc/s10052-022-10603-7} {\bibfield  {journal} {\bibinfo  {journal}
  {Eur. Phys. J. C}\ }\textbf {\bibinfo {volume} {82}},\ \bibinfo {pages} {633}
  (\bibinfo {year} {2022})},\ \Eprint {http://arxiv.org/abs/2106.08070}
  {arXiv:2106.08070 [gr-qc]} \BibitemShut {NoStop}%
\bibitem [{\citenamefont {Roman}(1993)}]{roman1993inflating}%
  \BibitemOpen
  \bibfield  {author} {\bibinfo {author} {\bibfnamefont {T.~A.}\ \bibnamefont
  {Roman}},\ }\href@noop {} {\bibfield  {journal} {\bibinfo  {journal}
  {Physical Review D}\ }\textbf {\bibinfo {volume} {47}},\ \bibinfo {pages}
  {1370} (\bibinfo {year} {1993})}\BibitemShut {NoStop}%
\bibitem [{\citenamefont {Kuo}\ and\ \citenamefont
  {Ford}(1993)}]{kuo1993semiclassical}%
  \BibitemOpen
  \bibfield  {author} {\bibinfo {author} {\bibfnamefont {C.-I.}\ \bibnamefont
  {Kuo}}\ and\ \bibinfo {author} {\bibfnamefont {L.}~\bibnamefont {Ford}},\
  }\href@noop {} {\bibfield  {journal} {\bibinfo  {journal} {Physical Review
  D}\ }\textbf {\bibinfo {volume} {47}},\ \bibinfo {pages} {4510} (\bibinfo
  {year} {1993})}\BibitemShut {NoStop}%
\bibitem [{\citenamefont {Farnes}(2018)}]{farnes2018unifying}%
  \BibitemOpen
  \bibfield  {author} {\bibinfo {author} {\bibfnamefont {J.~S.}\ \bibnamefont
  {Farnes}},\ }\href@noop {} {\bibfield  {journal} {\bibinfo  {journal}
  {Astronomy \& Astrophysics}\ }\textbf {\bibinfo {volume} {620}},\ \bibinfo
  {pages} {A92} (\bibinfo {year} {2018})}\BibitemShut {NoStop}%
\bibitem [{\citenamefont {Bekenstein}(1974)}]{bekenstein1974the}%
  \BibitemOpen
  \bibfield  {author} {\bibinfo {author} {\bibfnamefont {J.}~\bibnamefont
  {Bekenstein}},\ }\href@noop {} {\bibfield  {journal} {\bibinfo  {journal}
  {Lettere al Nuovo Cimento}\ }\textbf {\bibinfo {volume} {11}},\ \bibinfo
  {pages} {467} (\bibinfo {year} {1974})}\BibitemShut {NoStop}%
\bibitem [{\citenamefont {Carr}\ \emph {et~al.}(2015)\citenamefont {Carr},
  \citenamefont {Mureika},\ and\ \citenamefont {Nicolini}}]{Carr:2015nqa}%
  \BibitemOpen
  \bibfield  {author} {\bibinfo {author} {\bibfnamefont {B.~J.}\ \bibnamefont
  {Carr}}, \bibinfo {author} {\bibfnamefont {J.}~\bibnamefont {Mureika}}, \
  and\ \bibinfo {author} {\bibfnamefont {P.}~\bibnamefont {Nicolini}},\ }\href
  {\doibase 10.1007/JHEP07(2015)052} {\bibfield  {journal} {\bibinfo  {journal}
  {JHEP}\ }\textbf {\bibinfo {volume} {07}},\ \bibinfo {pages} {052} (\bibinfo
  {year} {2015})},\ \Eprint {http://arxiv.org/abs/1504.07637} {arXiv:1504.07637
  [gr-qc]} \BibitemShut {NoStop}%
\bibitem [{\citenamefont {Ali}(2022)}]{Ali:2022jna}%
  \BibitemOpen
  \bibfield  {author} {\bibinfo {author} {\bibfnamefont {A.~F.}\ \bibnamefont
  {Ali}},\ }\href@noop {} {\  (\bibinfo {year} {2022})},\ \Eprint
  {http://arxiv.org/abs/2210.13974} {arXiv:2210.13974 [quant-ph]} \BibitemShut
  {NoStop}%
\bibitem [{\citenamefont {Jafferis}\ \emph
  {et~al.}(2022{\natexlab{b}})\citenamefont {Jafferis}, \citenamefont
  {Zlokapa}, \citenamefont {Lykken}, \citenamefont {Kolchmeyer}, \citenamefont
  {Davis}, \citenamefont {Lauk}, \citenamefont {Neven},\ and\ \citenamefont
  {Spiropulu}}]{Jafferis:2022crx}%
  \BibitemOpen
  \bibfield  {author} {\bibinfo {author} {\bibfnamefont {D.}~\bibnamefont
  {Jafferis}}, \bibinfo {author} {\bibfnamefont {A.}~\bibnamefont {Zlokapa}},
  \bibinfo {author} {\bibfnamefont {J.~D.}\ \bibnamefont {Lykken}}, \bibinfo
  {author} {\bibfnamefont {D.~K.}\ \bibnamefont {Kolchmeyer}}, \bibinfo
  {author} {\bibfnamefont {S.~I.}\ \bibnamefont {Davis}}, \bibinfo {author}
  {\bibfnamefont {N.}~\bibnamefont {Lauk}}, \bibinfo {author} {\bibfnamefont
  {H.}~\bibnamefont {Neven}}, \ and\ \bibinfo {author} {\bibfnamefont
  {M.}~\bibnamefont {Spiropulu}},\ }\href {\doibase 10.1038/s41586-022-05424-3}
  {\bibfield  {journal} {\bibinfo  {journal} {Nature}\ }\textbf {\bibinfo
  {volume} {612}},\ \bibinfo {pages} {51} (\bibinfo {year}
  {2022}{\natexlab{b}})}\BibitemShut {NoStop}%
\bibitem [{\citenamefont {Ali}\ \emph {et~al.}(2022{\natexlab{b}})\citenamefont
  {Ali}, \citenamefont {Elmashad},\ and\ \citenamefont
  {Mureika}}]{Ali:2022ckm}%
  \BibitemOpen
  \bibfield  {author} {\bibinfo {author} {\bibfnamefont {A.~F.}\ \bibnamefont
  {Ali}}, \bibinfo {author} {\bibfnamefont {I.}~\bibnamefont {Elmashad}}, \
  and\ \bibinfo {author} {\bibfnamefont {J.}~\bibnamefont {Mureika}},\ }\href
  {\doibase 10.1016/j.physletb.2022.137182} {\bibfield  {journal} {\bibinfo
  {journal} {Phys. Lett. B}\ }\textbf {\bibinfo {volume} {831}},\ \bibinfo
  {pages} {137182} (\bibinfo {year} {2022}{\natexlab{b}})},\ \Eprint
  {http://arxiv.org/abs/2205.14009} {arXiv:2205.14009 [physics.gen-ph]}
  \BibitemShut {NoStop}%
\bibitem [{\citenamefont {Bosso}\ \emph {et~al.}(2022)\citenamefont {Bosso},
  \citenamefont {Petruzziello},\ and\ \citenamefont {Wagner}}]{Bosso:2022vlz}%
  \BibitemOpen
  \bibfield  {author} {\bibinfo {author} {\bibfnamefont {P.}~\bibnamefont
  {Bosso}}, \bibinfo {author} {\bibfnamefont {L.}~\bibnamefont {Petruzziello}},
  \ and\ \bibinfo {author} {\bibfnamefont {F.}~\bibnamefont {Wagner}},\ }\href
  {\doibase 10.1016/j.physletb.2022.137415} {\bibfield  {journal} {\bibinfo
  {journal} {Phys. Lett. B}\ }\textbf {\bibinfo {volume} {834}},\ \bibinfo
  {pages} {137415} (\bibinfo {year} {2022})},\ \Eprint
  {http://arxiv.org/abs/2206.05064} {arXiv:2206.05064 [gr-qc]} \BibitemShut
  {NoStop}%
\bibitem [{\citenamefont {Roman}(2006)}]{roman2006some}%
  \BibitemOpen
  \bibfield  {author} {\bibinfo {author} {\bibfnamefont {T.~A.}\ \bibnamefont
  {Roman}},\ }\enquote {\bibinfo {title} {Some thoughts on energy conditions
  and wormholes},}\ in\ \href {\doibase 10.1142/9789812704030_0236} {\emph
  {\bibinfo {booktitle} {The Tenth Marcel Grossmann Meeting}}}\ (\bibinfo
  {publisher} {World Scientific},\ \bibinfo {year} {2006})\ pp.\ \bibinfo
  {pages} {1909--1924}\BibitemShut {NoStop}%
\bibitem [{\citenamefont {Gaete}\ \emph {et~al.}(2022)\citenamefont {Gaete},
  \citenamefont {Jusufi},\ and\ \citenamefont {Nicolini}}]{Gaete:2022ukm}%
  \BibitemOpen
  \bibfield  {author} {\bibinfo {author} {\bibfnamefont {P.}~\bibnamefont
  {Gaete}}, \bibinfo {author} {\bibfnamefont {K.}~\bibnamefont {Jusufi}}, \
  and\ \bibinfo {author} {\bibfnamefont {P.}~\bibnamefont {Nicolini}},\ }\href
  {\doibase 10.1016/j.physletb.2022.137546} {\bibfield  {journal} {\bibinfo
  {journal} {Phys. Lett. B}\ }\textbf {\bibinfo {volume} {835}},\ \bibinfo
  {pages} {137546} (\bibinfo {year} {2022})},\ \Eprint
  {http://arxiv.org/abs/2205.15441} {arXiv:2205.15441 [hep-th]} \BibitemShut
  {NoStop}%
\bibitem [{\citenamefont {Garattini}(2019)}]{Garattini:2019ivd}%
  \BibitemOpen
  \bibfield  {author} {\bibinfo {author} {\bibfnamefont {R.}~\bibnamefont
  {Garattini}},\ }\href {\doibase 10.1140/epjc/s10052-019-7468-y} {\bibfield
  {journal} {\bibinfo  {journal} {Eur. Phys. J. C}\ }\textbf {\bibinfo {volume}
  {79}},\ \bibinfo {pages} {951} (\bibinfo {year} {2019})},\ \Eprint
  {http://arxiv.org/abs/1907.03623} {arXiv:1907.03623 [gr-qc]} \BibitemShut
  {NoStop}%
\bibitem [{\citenamefont {Isi}\ \emph {et~al.}(2013)\citenamefont {Isi},
  \citenamefont {Mureika},\ and\ \citenamefont {Nicolini}}]{Isi:2013cxa}%
  \BibitemOpen
  \bibfield  {author} {\bibinfo {author} {\bibfnamefont {M.}~\bibnamefont
  {Isi}}, \bibinfo {author} {\bibfnamefont {J.}~\bibnamefont {Mureika}}, \ and\
  \bibinfo {author} {\bibfnamefont {P.}~\bibnamefont {Nicolini}},\ }\href
  {\doibase 10.1007/JHEP11(2013)139} {\bibfield  {journal} {\bibinfo  {journal}
  {JHEP}\ }\textbf {\bibinfo {volume} {11}},\ \bibinfo {pages} {139} (\bibinfo
  {year} {2013})},\ \Eprint {http://arxiv.org/abs/1310.8153} {arXiv:1310.8153
  [hep-th]} \BibitemShut {NoStop}%
\bibitem [{\citenamefont {Carr}\ and\ \citenamefont
  {Hawking}(1974)}]{carr1974black}%
  \BibitemOpen
  \bibfield  {author} {\bibinfo {author} {\bibfnamefont {B.~J.}\ \bibnamefont
  {Carr}}\ and\ \bibinfo {author} {\bibfnamefont {S.~W.}\ \bibnamefont
  {Hawking}},\ }\href@noop {} {\bibfield  {journal} {\bibinfo  {journal}
  {Monthly Notices of the Royal Astronomical Society}\ }\textbf {\bibinfo
  {volume} {168}},\ \bibinfo {pages} {399} (\bibinfo {year}
  {1974})}\BibitemShut {NoStop}%
\bibitem [{\citenamefont {Hawking}(1971)}]{hawking1971gravitationally}%
  \BibitemOpen
  \bibfield  {author} {\bibinfo {author} {\bibfnamefont {S.}~\bibnamefont
  {Hawking}},\ }\href@noop {} {\bibfield  {journal} {\bibinfo  {journal}
  {Monthly Notices of the Royal Astronomical Society}\ }\textbf {\bibinfo
  {volume} {152}},\ \bibinfo {pages} {75} (\bibinfo {year} {1971})}\BibitemShut
  {NoStop}%
\bibitem [{\citenamefont {Debergh}\ \emph {et~al.}(2018)\citenamefont
  {Debergh}, \citenamefont {Petit},\ and\ \citenamefont
  {D’Agostini}}]{debergh2018evidence}%
  \BibitemOpen
  \bibfield  {author} {\bibinfo {author} {\bibfnamefont {N.}~\bibnamefont
  {Debergh}}, \bibinfo {author} {\bibfnamefont {J.-P.}\ \bibnamefont {Petit}},
  \ and\ \bibinfo {author} {\bibfnamefont {G.}~\bibnamefont {D’Agostini}},\
  }\href@noop {} {\bibfield  {journal} {\bibinfo  {journal} {Journal of Physics
  Communications}\ }\textbf {\bibinfo {volume} {2}},\ \bibinfo {pages} {115012}
  (\bibinfo {year} {2018})}\BibitemShut {NoStop}%
\bibitem [{\citenamefont {Mathur}(2009)}]{Mathur:2008wi}%
  \BibitemOpen
  \bibfield  {author} {\bibinfo {author} {\bibfnamefont {S.~D.}\ \bibnamefont
  {Mathur}},\ }\href {\doibase 10.1007/978-3-540-88460-6_1} {\bibfield
  {journal} {\bibinfo  {journal} {Lect. Notes Phys.}\ }\textbf {\bibinfo
  {volume} {769}},\ \bibinfo {pages} {3} (\bibinfo {year} {2009})},\ \Eprint
  {http://arxiv.org/abs/0803.2030} {arXiv:0803.2030 [hep-th]} \BibitemShut
  {NoStop}%
\bibitem [{\citenamefont {Ali}(2021)}]{Ali:2021ela}%
  \BibitemOpen
  \bibfield  {author} {\bibinfo {author} {\bibfnamefont {A.~F.}\ \bibnamefont
  {Ali}},\ }\href {\doibase 10.1142/S0217751X21501372} {\bibfield  {journal}
  {\bibinfo  {journal} {Int. J. Mod. Phys. A}\ }\textbf {\bibinfo {volume}
  {36}},\ \bibinfo {pages} {2150137} (\bibinfo {year} {2021})},\ \Eprint
  {http://arxiv.org/abs/2107.04643} {arXiv:2107.04643 [physics.gen-ph]}
  \BibitemShut {NoStop}%
\bibitem [{\citenamefont {Jusufi}(2023)}]{Jusufi:2023pzt}%
  \BibitemOpen
  \bibfield  {author} {\bibinfo {author} {\bibfnamefont {K.}~\bibnamefont
  {Jusufi}},\ }\href {\doibase 10.3390/universe9010041} {\bibfield  {journal}
  {\bibinfo  {journal} {Universe}\ }\textbf {\bibinfo {volume} {9}},\ \bibinfo
  {pages} {41} (\bibinfo {year} {2023})},\ \Eprint
  {http://arxiv.org/abs/2301.03590} {arXiv:2301.03590 [gr-qc]} \BibitemShut
  {NoStop}%
\bibitem [{\citenamefont {Almheiri}\ \emph {et~al.}(2020)\citenamefont
  {Almheiri}, \citenamefont {Hartman}, \citenamefont {Maldacena}, \citenamefont
  {Shaghoulian},\ and\ \citenamefont {Tajdini}}]{Almheiri:2019qdq}%
  \BibitemOpen
  \bibfield  {author} {\bibinfo {author} {\bibfnamefont {A.}~\bibnamefont
  {Almheiri}}, \bibinfo {author} {\bibfnamefont {T.}~\bibnamefont {Hartman}},
  \bibinfo {author} {\bibfnamefont {J.}~\bibnamefont {Maldacena}}, \bibinfo
  {author} {\bibfnamefont {E.}~\bibnamefont {Shaghoulian}}, \ and\ \bibinfo
  {author} {\bibfnamefont {A.}~\bibnamefont {Tajdini}},\ }\href {\doibase
  10.1007/JHEP05(2020)013} {\bibfield  {journal} {\bibinfo  {journal} {JHEP}\
  }\textbf {\bibinfo {volume} {05}},\ \bibinfo {pages} {013} (\bibinfo {year}
  {2020})},\ \Eprint {http://arxiv.org/abs/1911.12333} {arXiv:1911.12333
  [hep-th]} \BibitemShut {NoStop}%
\end{thebibliography}%

\end{document}